# Evidence for trap-assisted Auger recombination in MBE grown InGaN quantum wells by electron emission spectroscopy

Daniel J. Myers,[1,a)] Andrew C. Espenlaub,[1] Kristina Gelžinytė,[2] Erin C. Young,[1] Lucio Martinelli,[3] Jacques Peretti,[3] Claude Weisbuch,[1,3] and James S. Speck[1]

[1]Materials Department, University of California, Santa Barbara, California 93106, USA

[2]Institute of Applied Research, Vilnius University, Sauletekio 9-3, 10222 Vilnius, Lithuania

[3]Laboratoire de Physique de la Matière Condensée, CNRS, Ecole Polytechnique, IP Paris, 91128 Palaiseau Cedex, France

[a)]author to whom correspondence should be addressed: danielmyersj@gmail.com

ABSTRACT

We report on the direct measurement of hot electrons generated in the active region of blue light-emitting diodes grown by ammonia molecular beam epitaxy by electron emission spectroscopy. The external quantum efficiency of these devices is <1% and does not droop; thus, the efficiency losses from the intrinsic, interband, electron–electron–hole, or electron–hole–hole Auger should not be a significant source of hot carriers. The detection of hot electrons in this case suggests that an alternate hot electron generating process is occurring within these devices, likely a trap-assisted Auger recombination process.

Electron emission spectroscopy (EES) is a technique that offers a direct measurement of hot electron-generating efficiency loss mechanisms occurring in semiconductor devices.[1,2] In these measurements, the emitted electron energy distribution provides information about some of the relevant electron transport and carrier recombination mechanisms. This emission of electrically injected electrons into vacuum is made possible by taking advantage of a phenomenon known as negative electron affinity (NEA).[3]

EES measurements from GaN-based light-emitting diodes (LEDs) usually show four distinct peaks: two low-energy photoemission (PE) peaks due to light emitted by the LED photo-exciting electrons from the p-contact metals[4] and two higher energy peaks due to electrons that are emitted into vacuum directly from the semiconductor surface. The two semiconductor emitted peaks correspond to electrons emitted from a high energy conduction band side-valley (SV) as well as electrons emitted from the Γ-valley conduction band minimum. The SV peak is indicative of electrons that have undergone some hot electron generating process, typically electron–electron–hole (eeh) Auger recombination, and subsequently relaxed into the SV. The Γ-valley peak is indicative of electrons that have bypassed the QWs in the active region (Fig. 1). The energy separation between the SV and C valleys is roughly 1 eV, which has been confirmed by other methods.[5–7] Because the energy imparted to the hot electron in eeh Auger recombination for blue LEDs is roughly 2.75 eV, it is possible for these hot electrons to scatter in the higher energy SV and survive on the p-GaN surface of the device where these electrons are emitted and measured using a spectrometer.[1,2,4]

High efficiency (>70% EQE) InGaN/GaN LEDs are widely available today and are produced by metal organic chemical vapor deposition (MOCVD).[8] Another widely used semiconductor growth technique is molecular beam epitaxy (MBE), but InGaN/GaN LEDs produced by MBE typically have radiative efficiencies of only a few percent, despite excellent current–voltage (I–V) characteristics.[9] This low efficiency behavior of MBE LEDs is generally attributed to point defects or native impurities that act as non-radiative recombination centers. There have been studies to identify these potential impurities, but the low efficiency of MBE produced LEDs remains an open question.[10]

The non-radiative recombination mechanisms present in high efficiency InGaN/GaN LEDs have been widely discussed in the published literature.[11] At low carrier densities, the dominant efficiency loss mechanism is Shockley–Read–Hall (SRH) recombination.[12,13] At high carrier densities, it is generally accepted that interband Auger recombination is the dominant non-radiative recombination mechanism and is responsible for the phenomenon known as "efficiency droop."[1,2,4] Interband Auger recombination is a three-body carrier recombination process that involves either two electrons and one hole (eeh) or one electron and two holes (ehh).[18]

Studies of these mechanisms often rely on indirect measurements of carrier dynamics by analysis of LED electroluminescence efficiency. Most commonly, these studies make use of the "ABC" rate equation model for radiative efficiency.[14] This model requires several approximations. Namely, it is assumed that the electron and hole concentrations are equal. This is generally untrue for several reasons: first, polarization charges at heterointerfaces spatially separate electrons and holes; also, large differences in the electron and hole mobilities lead to non-uniform hole injection, and finally, natural alloy composition fluctuations within the InGaN quantum wells can localize carriers (both electrons and holes) to indium-rich regions,[15,16] leading to spatially inhomogeneous and, in general, unequal concentrations. However, with careful measurements and device design, the shortcomings of the ABC rate equation can be overcome and this model can be used to provide valuable information and insight into the dynamics and characteristics of the relevant recombination processes.

In this paper, we report on the direct measurements of the energy distribution of vacuum emitted electrons from low efficiency MBE LEDs. These data are assembled in electron energy distribution curves (EDCs) and shows that a significant portion of the emission current was due to hot carriers emitted from a high energy, conduction band SV. Normally, these hot carriers are a signature of eeh Auger recombination occurring in the LED active region, a process that should only be appreciable at high current densities.

In conjunction with the EES measurements, we quantify the electroluminescence efficiency of the low efficiency MBE LEDs and show that these devices exhibit no drop in efficiency as the current density is increased. The combination of EES measurements capturing the signatures of hot carrier generation, no efficiency droop, and very low efficiency suggests that there must be another mechanism capable of generating hot electrons occurring in these devices, which is responsible for the low efficiency of MBE grown LEDs.

Based on the data collected, we believe that the origin of these hot electrons in MBE LEDs is likely a trap-assisted Auger recombination (TAAR) process.[17–26] For electron–electron (ee) TAAR, capture of an electron from the conduction band directly into a trap state within the forbidden gap occurs concurrently with the generation of a second hot electron, instead of the emission of phonons as in SRH. This hot electron has sufficient energy to scatter into the high energy SV where it can be emitted into vacuum and measured by EES (shown schematically in Fig. 1). A similar TAAR mechanism occurring in AlGaN heterostructure barriers has recently been observed.[27]

The devices measured in this paper consist of LEDs grown by ammonia MBE. The epitaxial structure is as follows: 3 μm GaN on a sapphire template ∣ 450 nm n-GaN [Si] = $2.5 \times 10^{18}$ cm$^{-3}$ ∣ 10 nm UID GaN barrier ∣ 3 nm InGaN QW ∣ 12 nm UID GaN barrier ∣ 80 nm p-GaN [Mg] = $3 \times 10^{19}$ cm$^{-3}$ ∣ a 5 nm p$^{++}$ contact layer. The electroluminescence emitted by these LEDs was centered at 475 nm. Full details of the electrostatic energy analyzer and EES device design are reported elsewhere.[4]

The specific constraints required for EES devices prevented them from being useful for absolute measurement of the electroluminescence efficiency. Separate devices were fabricated, singulated, and encapsulated from the same epitaxial material that was used for the EES devices. The absolute EQE of these devices was then measured in an integrating sphere. The EQE of these devices reached a maximum value of ~0.2% assuming near unity injection efficiency.[28] Because of the low light output power of these low efficiency LEDs, the measurements in the integrating sphere had a low signal-to-noise ratio, even with exceptionally long integration times. Therefore, the maximum absolute EQE of ~0.2% was used to scale much less noisy measurements of the relative EQE performed on a third set of devices fabricated from the same MBE LED epitaxial material. These devices were left on-wafer, and their relative EQE was measured using a setup equipped with an actively temperature-controlled stage and a Si photodiode amplified by a FEMTO high-gain current amplifier (108 V/A). Both the geometry of the on-wafer LEDs and the measurement setup are described elsewhere.[29] The EQE measurements are described in the study by Espenlaub et al.[30] and are reproduced here (Fig. 2) to show the lack of efficiency droop in these MBE grown LEDs.

The EQE measurements show that over a wide range of carrier densities, the radiative efficiency remains constant and at a very low value (Fig. 2). To explain these data, we consider the different nonradiative recombination mechanisms that could contribute to this EQE behavior. Because radiative recombination has a quadratic dependence on the carrier density and SRH is linear, the LED efficiency would scale as $Bn/A$ and increase linearly with the carrier density. Thus, if SRH recombination was the dominant efficiency loss mechanism in MBE LEDs, we would expect that efficiency would increase with carrier density.

A similar argument can be made for interband Auger recombination. If interband Auger recombination was a significant non-radiative recombination mechanism occurring in this device, then efficiency droop should occur at sufficiently high carrier densities. The cubic dependence of interband Auger recombination would yield efficiency that scales as $B/Cn$ and would decrease with increasing current density.

In EES measurements, we directly measure hot electrons emitted from these low efficiency LEDs, typically a signature of interband Auger recombination.[1] This explanation is not applicable for the MBE LEDs as interband, eeh, and Auger recombination would cause efficiency to droop with increasing current density. The hot electrons emitted from these LEDs must be due to another hot carrier generating mechanism that has a similar dependence on carrier density as radiative recombination. We believe this process is TAAR.

To see this, a modification can be made to the, admittedly simplified, 'ABC' model discussed above, adding a term to the total recombination rate, $B'n^2$, to represent a non-radiative TAAR process, where n is the carrier density, and B' is the TAAR rate coefficient (note that, in contrast to B, B' is not an intrinsic parameter and will vary proportional to the concentration of TAAR inducing impurities). Now, the radiative efficiency in the QW is given by $Bn^2/(An + Bn^2 + B'n^2 + Cn^3)$, where A, B, and C are the SRH, radiative, and interband Auger coefficients, respectively. Choosing the values for the A, B, B', and C coefficients in the InGaN QW,[30] of $5.3 \times 10^8$ 1/s, $5.0 \times 10^{-12}$ cm$^3$/s, $5.1 \times 10^{-10}$ cm$^3$/s, and $6.4 \times 10^{-32}$ cm$^6$/s, respectively, and assuming that the light extraction efficiency in the singulated LEDs is ~34%, the EQE curve in Fig. 2 is well reproduced by this model and the lifetimes are consistent with the very short lifetimes (~10 ps) determined by time resolved photoluminescence measurements on similarly prepared MBE materials from our group.[30] We note that the B and C coefficients from our fit closely agree with those for ABC fits that include escape of Auger generated carriers ("Quantra/Cels (hot carriers)" model in Table I in Ref. 31).

The EDCs we measured from the low efficiency MBE LEDs showed two distinct peaks at drive current densities as low as 8.75A/cm$^2$ (Fig. 3). The EDCs show a low energy peak, consistent with the expected energy position of electrons emitted from the Γ-valley conduction band minimum, and a high energy peak ~1 eV above the Γ-valley peak, the expected energy position of the SV peak, thus demonstrating that a hot electron generating process must be occurring within the device. Unlike previous studies on high efficiency devices,[1,4,27] the presence of an SV peak cannot be an indicator of interband Auger recombination due to the lack of an appreciable efficiency droop. We believe that TAAR is a strong candidate for the efficiency limiting mechanism in these devices because of the efficiency behavior of these devices in conjunction with the direct measurements of hot carriers emitted from the p-type surface indicating some hot carrier generating mechanism is occurring. Furthermore, the abundance of point defects and impurities in MBE grown III-N LEDs could potentially provide ample amounts of nonradiative sites for TAAR to occur.

The relationship between the diode current density and the hot electron generation rate—perhaps by TAAR—can be extracted from the EES data by integrating the area under the SV peaks in the EDCs (that is proportional to the hot electron generation rate). The peaks were fit with exponentially modified Gaussian (EMG) functions by non-linear least squares; this peak shape was chosen due to the specifics of the EES measurement technique. The integrated SV peak area is plotted in Fig. 4. The error bars in this figure represent the estimate of the error in the peak area, derived from the residuals in the non-linear least squares fit; the larger errors at high drive currents are due to a peak shape more complicated than the assumed EMG shape. The dashed line shows the expected relationship between the TAAR recombination current ($\propto B'n^2$), and the observed integrated SV peak area, as a function of total drive current. The model shows agreement with the EES data. The TAAR recombination current is calculated using the modified "ABB'C" model described above (with the parameter values in given in the text), and scaled as a whole to match the magnitude of the data (the EES measurements provide only a relative, and low efficiency, measure of the total hot carrier generation rate within the device). The slow turn-on of the TAAR component of the current in the model is due to the dominance of conventional SRH recombination at low drive currents; in the measured integrated SV peak areas this manifests as the apparent offset of the linear portion of the data to ~15mA.

To preserve charge neutrality, the total injected current must equal the total recombination current (the sum of all recombination processes). The proportionality of SRH, TAAR, radiative, and interband Auger recombination is shown in Fig. 5, using the rate coefficients discussed earlier. The spectroscopic measurement of hot electrons emitted from the LED surface provides unambiguous detection of a hot carrier generating process occurring in these devices. The recombination mechanism generating these hot electrons does so in a regime where interband Auger recombination remains less than 0.1% over the entire range of current densities measured.

In conclusion, we have determined that the poor efficiency characteristics of MBE LEDs are linked to a mechanism that generation of hot electrons, likely an Auger-type process, such as TAAR, occurs in InGaN quantum wells. This is evidenced by the presence of hot electrons measured directly by electron emission spectroscopy, in the absence of an efficiency droop (usually due to interband Auger). The integrated area under the SV peaks (proportional to the hot electron generation rate within the device) scales linearly with the total current, as would be expected in very low efficiency devices where the dominant recombination mechanism is a hot electron generating process such as TAAR.

The authors would like to acknowledge Wan Ying Ho for her contributions to this work. This work was supported by the U.S. Department of Energy under the Office of Energy Efficiency & Renewable Energy (EERE) Award No. DE-EE0007096. A portion of work was completed at the UCSB Nanofabrication Facility.

**REFERENCES**

1. J. Iveland, L. Martinelli, J. Peretti, J. S. Speck, and C. Weisbuch, Phys. Rev. Lett. 110, 177406 (2013).

2. J. Iveland, M. Piccardo, L. Martinelli, J. Peretti, J. W. Choi, N. Young, S. Nakamura, J. S. Speck, and C. Weisbuch, Appl. Phys. Lett. 105, 052103 (2014).

3. C. I. Wu and A. Kahn, Appl. Surf. Sci. 162, 250 (2000).

4. D. J. Myers, K. Gelžinytė, W. Y. Ho, J. Iveland, L. Martinelli, J. Peretti, C. Weisbuch, and J. S. Speck, J. Appl. Phys. 124, 055703 (2018).

5. M. Piccardo, L. Martinelli, J. Iveland, N. Young, S. P. Denbaars, S. Nakamura, J. S. Speck, C. Weisbuch, and J. Peretti, Phys. Rev. B 89, 235124 (2014).

6. S. Marcinkevičius, T. K. Uždavinys, H. M. Foronda, D. A. Cohen, C. Weisbuch, and J. S. Speck, Phys. Rev. B 94, 235205 (2016).

7. S. Wu, P. Geiser, J. Jun, J. Karpinski, D. Wang, and R. Sobolewski, J. Appl. Phys. 101, 043701 (2007).

8. S. Nakamura, T. Mukai, and M. Senoh, Appl. Phys. Lett. 64, 1687 (1994).

9. K. Johnson, V. Bousquet, S. E. Hooper, M. Kauer, C. Zellweger, and J. Heffernan, Electr. Lett. 40, 1299 (2004).

10. E. C. Young, N. Grandjean, T. E. Mates, and J. S. Speck, Appl. Phys. Lett. 109, 212103 (2016).

11. D. Feezell and S. Nakamura, C. R. Phys. 19, 113 (2018).

12. A. Alkauskas, Q. Yan, and C. G. de Walle, Phys. Rev. B 90, 75202 (2014).

13. A. Alkauskas, C. E. Dreyer, J. L. Lyons, and C. G. Van De Walle, Phys. Rev. B 93, 201304(R) (2016).

14. Y. C. Shen, G. O. Mueller, S. Watanabe, N. F. Gardner, A. Munkholm, and M. R. Krames, Appl. Phys. Lett. 91, 141101 (2007).

15. Y. R. Wu, R. Shivaraman, K. C. Wang, and J. S. Speck, Appl. Phys. Lett. 101, 083505 (2012).

16. T. J. Yang, R. Shivaraman, J. S. Speck, and Y. R. Wu, J. Appl. Phys. 116, 113104 (2014).

17. P. T. Landsberg, D. A. Evans, and C. Rhys-Roberts, Proc. Phys. Soc. 83, 325 (1964).

18. P. T. Landsberg and D. J. Robbins, Solid State Electron. 21, 1289 (1978).

19. D. J. Robbins, P. T. Landsberg, D. M. Eagles, and C. J. Hearn, Proc. Phys. Soc. 84, 915 (1964).

20. A. Haug, Phys. Status Solidi 97, 481 (1980).

21. A. Haug, Phys. Status Solidi 108, 443 (1981).

22. D. J. Robbins, J. Phys. C Solid State Phys. 16, 3825 (1983).

23. V. B. Khalfin, M. V. Strikha, and I. N. Yassievich, Phys. Status Solidi 132, 203 (1985).

24. A. Hangleiter, Phys. Rev. B 37, 2594 (1988).

25. A. Haug, Appl. Phys. A Solids Surf. 56, 567 (1993).

26. A. Hangleiter, Phys. Rev. Lett. 55, 2976 (1985).

27. D. J. Myers, K. Gelžinytė, A. I. Alhassan, L. Martinelli, J. Peretti, S. Nakamura, C. Weisbuch, and J. S. Speck, Phys. Rev. B 100, 125303 (2019).

28. N. F. Gardner, G. O. Müller, Y. C. Shen, G. Chen, S. Watanabe, W. Götz, and M. R. Krames, Appl. Phys. Lett. 91, 243506 (2007).

29. A. C. Espenlaub, A. I. Alhassan, S. Nakamura, C. Weisbuch, and J. S. Speck, Appl. Phys. Lett. 112, 141106 (2018).

30. A. C. Espenlaub, D. J. Myers, E. C. Young, S. Marcinkevičius, C. Weisbuch, and J. S. Speck, J. Appl. Phys. 126, 184502 (2019).

31. J. Piprek, F. Römer, and B. Witzigmann, Appl. Phys. Lett. 106, 101101 (2015).

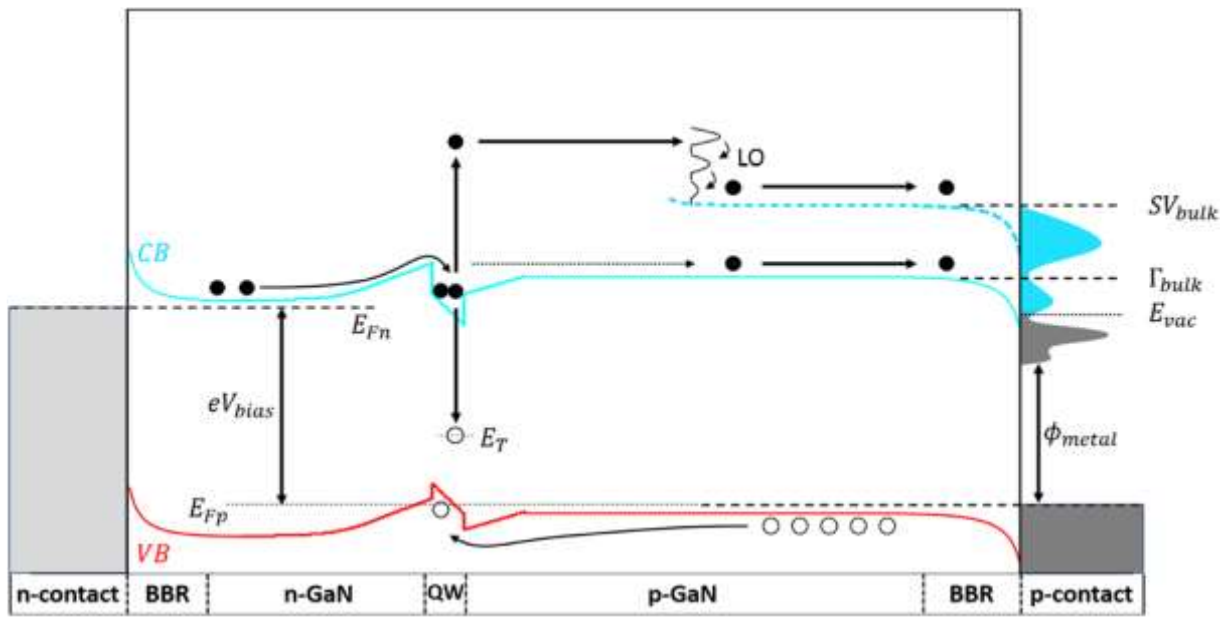

FIG. 1. Depiction of the TAAR mechanism occurring in an SQW LED and the relevant energy levels and transport to the surface in the SV.

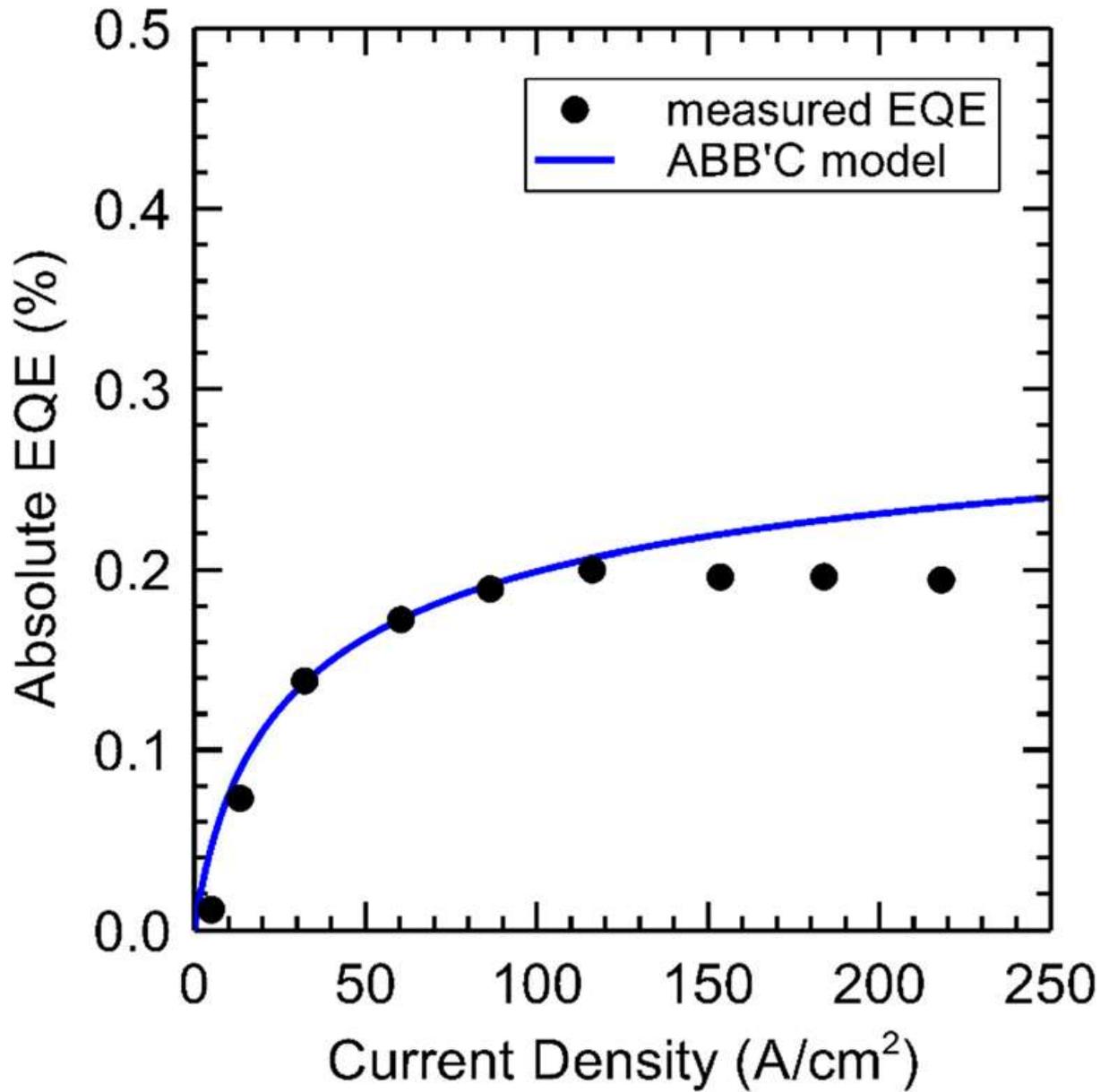

FIG. 2. The relative EQE of the separate, on-wafer MBE LEDs fabricated from the same epitaxial material as the EES devices, scaled by the measured peak absolute EQE value (~0.2%), (symbols). The model agrees well with the observed droopless EQE data.

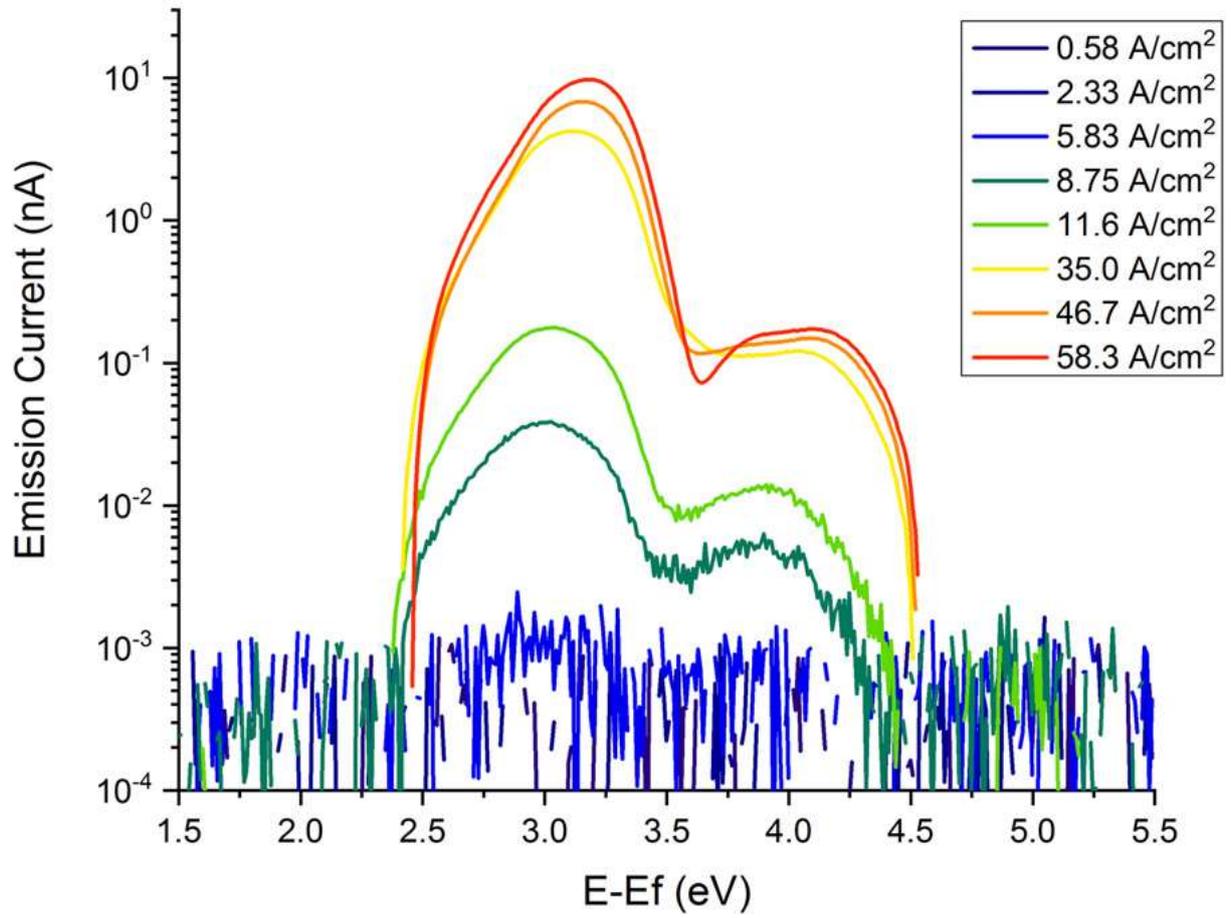

FIG. 3. EDCs from MBE LEDs at varying current densities. The low energy peak is due to electrons emitted from the Γ-valley conduction band minimum, while the higher energy peak is likely due to TAAR.

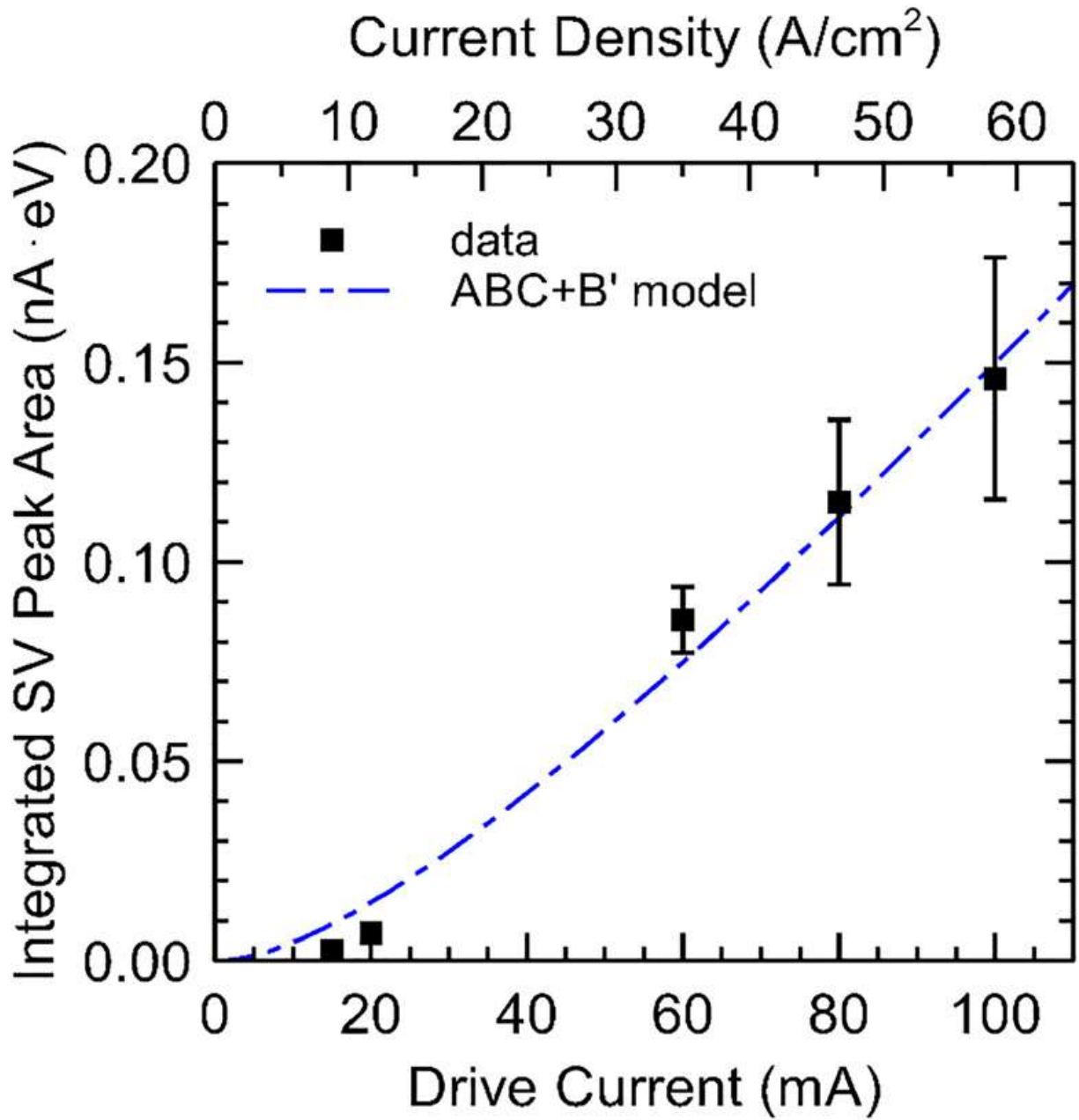

FIG. 4. Integrated SV peak intensity at varying diode currents (symbols). The linear relationship between the SV peak and the current density, above ~15 mA drive current, is consistent with the droopless EQE. The TAAR component of the total current density (dashed line) was calculated using the modified 'ABB'C' model and scaled to match the magnitude of the data.

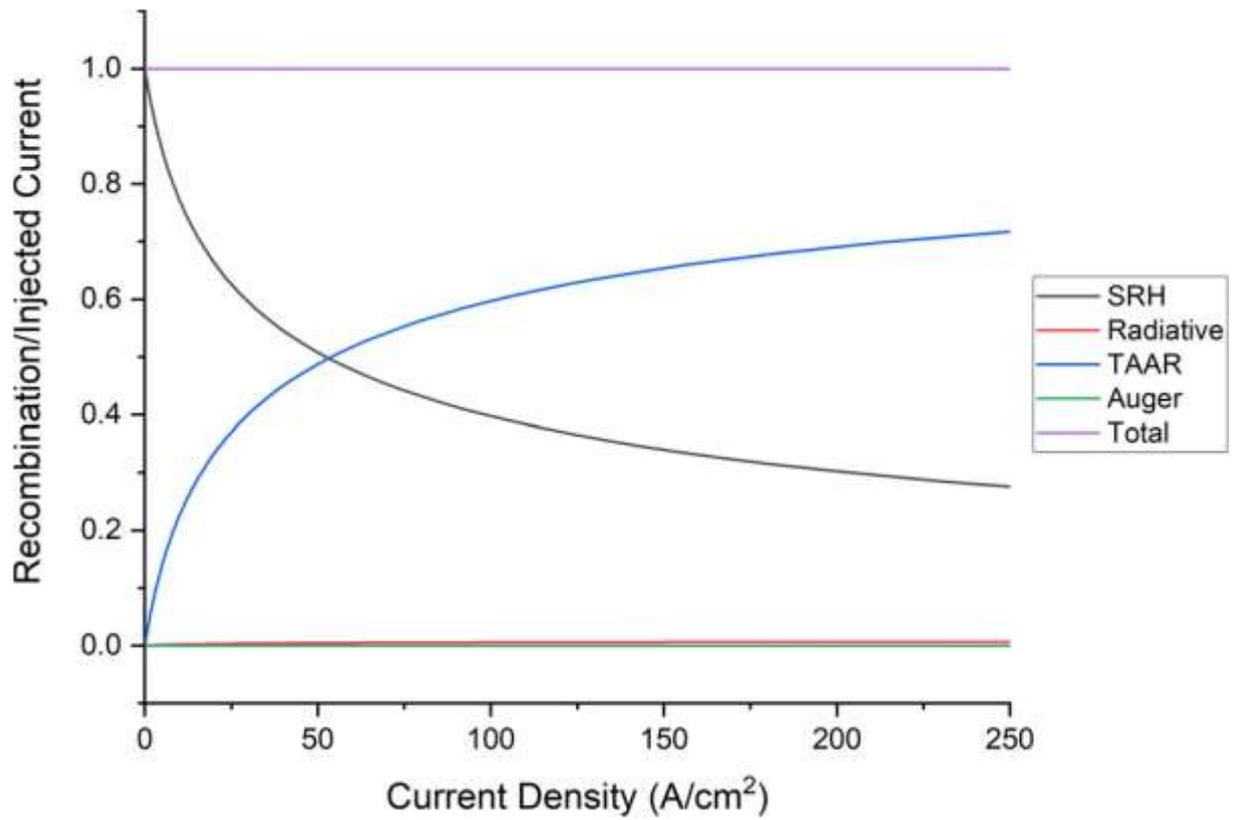

FIG. 5. Normalized plot of the various recombination currents, as a function of injected current density. At 250 A/cm$^2$, interband Auger recombination makes up about 0.025% ($\approx$0.06 A/cm$^2$) of the total recombination current. This is negligible when compared to the effects of a TAAR like process ($\approx$72%, 180 A/cm$^2$).